\documentstyle[12pt,aaspp4,flushrt]{article}

\newcommand\axpa{1RXS~J170849.0$-$400910}
\newcommand\axpb{1E~2259+586}

\newcommand\rxte{{\it RXTE}}

\def\lapp{\ifmmode\stackrel{<}{_{\sim}}\else$\stackrel{<}{_{\sim}}$\fi}
\def\gapp{\ifmmode\stackrel{>}{_{\sim}}\else$\stackrel{>}{_{\sim}}$\fi}

\begin{document}

\title{Precision Timing of Two Anomalous X-Ray Pulsars}

\author{Victoria M. Kaspi\altaffilmark{1}, Deepto Chakrabarty and Julia Steinberger}
\affil{Department of Physics and Center for Space Research, Massachusetts Institute of Technology, 70 Vassar Street, Cambridge, MA 02139}

\altaffiltext{1}{Alfred P. Sloan Research Fellow}

\medskip
\centerline{To appear in {\sc The Astrophysical Journal Letters}}

\begin{abstract}

We report on long-term X-ray timing of two anomalous X-ray pulsars,
\linebreak \axpa\ and \axpb, using the {\it Rossi X-ray Timing Explorer.}
In monthly observations made over 1.4~yr and 2.6~yr for the two pulsars,
respectively, we have obtained phase-coherent timing solutions which imply
that these objects have been rotating with great stability throughout
the course of our observations.  For \axpa, we find a rotation frequency of
0.0909169331(5) Hz and frequency derivative $-15.687(4) \times
10^{-14}$~Hz~s$^{-1}$, for epoch MJD 51215.931.  For \axpb, we find
a rotation frequency of 0.1432880613(2)~Hz, and frequency derivative
$-1.0026(7) \times 10^{-14}$~Hz~s$^{-1}$, for epoch MJD 51195.583.  RMS
phase residuals from these simple models are only $\sim$0.01 cycles for both
sources.  We show that the frequency derivative for \axpb\ is inconsistent
with that inferred from incoherent frequency observations made over the last
20~yr.  Our observations are consistent with the magnetar hypothesis and
make binary accretion scenarios appear unlikely.

\end{abstract}

\keywords{stars: neutron --- Pulsars: individual (\axpa, \axpb) --- X-rays: stars}

\section{Introduction}
\label{sec:intro}

The nature of anomalous X-ray pulsars (AXPs) has been a mystery since
the discovery of the first example (\axpb) nearly 20 years ago
(\cite{fg81}).  The properties of AXPs can be summarized as follows
(see also \cite{ms95,vtv95,gv98}): they exhibit X-ray pulsations in
the range $\sim$5--12~s; they have pulsed X-ray luminosities in the
range $\sim 10^{34}-10^{35}$~erg~s$^{-1}$; they spin down regularly
within the limited timing observations available, with some
exceptions; their X-ray luminosities are much greater than the rate of
loss of rotational kinetic energy inferred from the observed
spin-down; they have spectra that are characterized by thermal
emission with $kT \sim 0.4$~keV, with evidence for a hard tail in some
sources; they are found in the plane of the Galaxy; and two of the six
certain members of the class appear to be located at the geometric
centers of apparent supernova remnants (\cite{fg81,vg97}).  Soft gamma
repeaters also exhibit AXP-like pulsations in quiescence
(e.g. \cite{kds+98}).

Mereghetti \& Stella (1995) suggested \nocite{ms95} that AXPs are accreting
from a low mass companion.  However increasingly
this model has become difficult to reconcile with observations.  The
absence of Doppler shifts even on short time scales
(e.g. \cite{mis98}), the absence of a detectable optical/IR companion
or accretion disk (see \cite{ms95}), the apparent associations with
supernova remnants, the apparent steady spin down within the limits of
current observations (e.g. \cite{gvd99}), and AXP spectra that are very
different from those of known accreting sources, all argue against
an accretion origin for the X-rays.

Recently, it has been argued that the AXPs are young, isolated, highly
magnetized neutron stars or ``magnetars'' (\cite{td96a,hh97}).
Evidence for this is primarily the inferred strength of the surface
dipolar magnetic field required to slow the pulsar down {\it in
vacuo}: $\sim 10^{14}-10^{15}$~G.  The spin-down ages in this model,
inferred assuming small birth spin periods, are in the range of
$\sim$8--200~kyr.  This suggested youth is supported by the two
apparent supernova remnant associations.  Additional circumstantial
supporting evidence comes from AXPs location close to the Galactic
plane, consistent with their being isolated neutron stars near their
birth place, as well as from interpreting apparent deviations from
spin-down as glitches (\cite{hh99}) similar to those seen in radio
pulsars (e.g. \cite{kmj+92}).  Recently, deviations from simple
spin-down have been suggested, under the magnetar hypothesis, to be
due to radiative precession, originating in the asphericity of the
neutron star produced by the strong magnetic field (\cite{mel99}).

One way to test both the magnetar and accretion models is through
timing observations.  The spin down of some AXPs has been monitored by
considering the measured frequency at individual epochs
(e.g. \cite{bsss98,gvd99}).  However those measurements have been
sparse and are only marginally sensitive to spin irregularities on
time scales of weeks to months, relevant to glitches or accretion
torque fluctuations.  Further, high-precision determination of the
spin evolution over a long baseline is necessary to look for ``timing
noise'' as is seen in many young radio pulsars
(e.g. \cite{antt94,kms+94,lyn96}), to obtain a reliable measurement of
a braking index, and to search for precession (\cite{mel99}).  Whether
such high precision is possible to achieve with AXP timing has not,
until now, been established.

Here we report on X-ray monitoring observations made with the {\it
Rossi X-ray Timing Explorer} ({\em RXTE}) in which, for the first
time, we determine high-precision spin parameters using long-term
phase-coherent timing of two AXPs.  The sources, \axpa\
(\cite{snt+97,ics+99}) and \axpb\ (\cite{bs96,pof+98,bsss98}) have
periods of 11~s and 7~s, respectively.  The \rxte\ AXP monitoring
project is part of a larger effort to time coherently several AXPs.
Results for other sources will be presented elsewhere.

\section{Observations and Results}
\label{sec:obs}

Our observations were made using the {\em RXTE} Proportional Counter
Array (PCA) (\cite{jsss+96}).  The detector consists of five identical
multi-anode proportional counter units (PCUs) each containing a front
propane anticoincidence layer followed by several xenon/methane
layers.  The detector operates in the 2--60 keV range, with a total
effective area of $\sim$6500 cm$^2$ and a 1$^\circ$ field of view.  In
addition to the standard data modes, data were collected in the {\tt
GoodXenonwithPropane} mode, which records the arrival time (1 $\mu$s
resolution) and energy (256-channel resolution) of every unrejected
xenon event as well as all of the propane layer events.  To maximize
the sensitivity to the targets, which have soft spectra, we restricted
the analysis to unrejected events in the top xenon layer of each PCU
and chose an optimal energy range for each source: absolute
channels\footnote{These channel-to-energy conversions are for {\em
RXTE} gain epoch 3, from 1996 April 15 to 1999 March 22, averaged over
the five PCUs.}  6--14 (2.5--5.4 keV) for \axpa\ and absolute channels
6--24 (2.5--9.1 keV) for \axpb.  The observations were reduced using
MIT-developed software for handling raw spacecraft telemetry packet
data.  Data from the different PCUs were merged and binned at 62.5~ms
and 31.25~ms resolution for \axpa\ and \axpb, respectively.  The data
were then reduced to Barycentric Dynamical Time (TDB) at the solar
system barycenter using the source positions in Table 1 and the JPL
DE200 solar system ephemeris (\cite{snw+92}) and stored on disk as one
time series per observing epoch.

Our strategy for attempting phase-coherent timing of these
sources made use of standard radio pulsar techniques.  Pulse phase at
any time $t$ can be expressed as
\begin{equation}
\phi(t) = \phi(t_0) + \nu (t-t_0) + \frac{1}{2} \dot{\nu} (t-t_0)^2 + ...,
\end{equation}
where $t_0$ is a reference epoch, and $\nu$ and $\dot{\nu}$ are the 
spin frequency and its time derivative.  Phase-coherent timing
amounts to counting all pulses over the entire observing span.
To achieve this, uncertainties in the first-gues spin parameters $\nu$ and
$\dot{\nu}$ must be sufficiently small that the discrepancy between 
observed and predicted arrival time differ by only a fraction of 
the period.  To achieve this goal,
we observed each source at two closely spaced
epochs (i.e. within one day), then at a third epoch several days later.  This
spacing was chosen to determine an initial $\nu$, by absolute pulse
numbering, of sufficient precision to predict $\phi$ for the next
observation roughly one month later.  Subsequent monitoring was
done at roughly one month intervals.  Once phase
connection was achieved with the $\sim$6~months of monitoring data,
we also included public \rxte\ archival data.
For \axpa, the total data set consists of 19 arrival times obtained between
1998 January 13 and 1999 May 26.  For \axpb, we have 33
arrival times obtained between 1996 September 29 and 1999 May 12.

Our procedure included the following steps.  The first barycentered,
binned time series for the closely spaced set of three observations
were folded at the best-estimate period determined via Fourier
transform, using a unique reference epoch.  The folded pulse profiles
were cross-correlated with a high-signal-to-noise template in the
Fourier domain and the phase offsets were recorded.  We suppressed
high-order harmonics in the pulse profile using a frequency-domain
filter to avoid contamination by bin-to-bin Poisson fluctuations.  A
precise $\nu$ was then determined by demanding that an integer number
of pulses occur between each observation.  Profiles were then
re-folded, still with respect to a fixed epoch, using the improved
$\nu$.  After each new observation, this process was repeated, also
including the effect of $\dot{\nu}$.  Phase residuals were examined to
verify that there were no missed pulses, then fit with a quadratic
function to determine the optimal $\nu$ and $\dot{\nu}$.
Uncertainties on measured pulse phases were determined using Monte
Carlo simulations.  We verified this procedure and its results by
extracting absolute average pulse arrival times in TDB at the solar
system barycenter from the optimally folded profiles, and using the
{\tt TEMPO} pulsar timing software package ({\tt
http://pulsar.princeton.edu/tempo}), in common use in radio pulsar
timing.

Best fit $\nu$ and $\dot{\nu}$ for each source are given in
Table~\ref{ta:parms}.  These values were measured with $\ddot{\nu}$
fixed at zero.  Corresponding arrival time residuals are shown in
Figures~1 and 2.  In both cases, the RMS residual is $\sim$0.01$P$,
where $P=1/\nu$.  We also tried fitting for $\ddot{\nu}$; the results
are given in Table~\ref{ta:parms}.  For \axpb, the fitted
$\ddot{\nu}$ is consistent with zero; we provide a $3\sigma$ upper
limit.  For \axpa, the fitted $\ddot{\nu}$ is
marginally significant at the 4$\sigma$ level; however, a fit omitting
only the first point reduces the significance to 2.6$\sigma$.
We therefore quote the current best-fit value in parentheses only;
further timing observations will decide if the observed $\ddot{\nu}$
is truly significant.

\section{Discussion}
\label{sec:disc}

Using a very simple spin-down model, we have maintained phase
coherence for\linebreak \axpa\ and \axpb\ with phase residuals of only
$\sim$1\%, comparable to or smaller than those measured for most radio
pulsars (e.g. \cite{antt94}).  This demonstrates that these AXPs are
extremely stable rotators.  This stability is consistent with the
magnetar hypothesis because isolated rotating neutron stars are
expected to spin-down with great regularity, as is seen in the radio
pulsar population.

We can compare our pulse ephemerides with past period measurements to
see whether there have been a deviation from a simple spin-down law.
For \axpa, only two previous period measurements have been reported
(\cite{snt+97,ics+99}).  The pulse parameters listed in
Table~\ref{ta:parms}, extrapolated to the epochs of the previous
observations, agree with the published values within uncertainties.
Thus, the spin-down has been regular for at least 1.4 yr
prior to the commencement of our observations.

For \axpb, spin frequencies have been measured occasionally since 1978
(see Baykal \& Swank 1996 and references therein).\nocite{bs96}
Figure~\ref{fig:freq2259} shows the differences between previously
measured spin frequencies and those predicted by our timing ephemeris
(Table~\ref{ta:parms}).  The error bars represent the published one
standard deviation measurement uncertainties.  Our measured
$\dot{\nu}$ is not consistent with the long-term $\dot{\nu}$: all
observed frequencies were significantly larger than predicted by the
extrapolation of the current $\nu$ and $\dot{\nu}$.  A least squares
fit to the data shown in Figure~\ref{fig:freq2259} gives $\Delta
\dot{\nu} = 1.328(9) \times 10^{-15}$~Hz~s$^{-1}$, though the linear
fit is poor because of apparent short-time-scale fluctuations.  Thus,
the current value of $\dot{\nu}$, measured over the past 2.6~yr, is
smaller than that of the long-term trend by $\sim 10$\%.

Melatos (1999) suggests that the large magnetic field inferred in the
magnetar model should result in significant deviations from sphericity of
the neutron star, with the principle axis misaligned with the spin axis.
Under such circumstances, the star undergoes radiative precession with
period $\sim$10~yr.  Given the epoch and duration of our observations of
\axpb, the manifestation of such precession is completely covariant with
$\dot{\nu}$.  However, the implied deviation of $\dot{\nu}$ from the
long-term trend is consistent with the Melatos (1999) prediction, though
smaller by a factor $\sim$3.5.  An unambiguous test of this model can be
provided by periodic timing residuals on a time scale of a decade.
Note that the change in $\dot{\nu}$ that we observe for \axpb\ relative to
the long-term trend is not consistent with those observed after glitches in
radio pulsars, in which the absolute magnitude of the post-glitch
$\dot{\nu}$ is larger than the pre-glitch value (\cite{sl96}).

Radio pulsars, especially young ones, exhibit deviations from simple
spin-down laws which appear to be a result of random processes
(\cite{ch80}).  These deviations are not physically understood and are
commonly referred to as ``timing noise'' (see \cite{lyn96} for a
review).  The measured $\ddot{\nu}$'s for \axpa\ and \axpb\
(Table~\ref{ta:parms}) can be compared with those of radio pulsars.
Noise level has been quantified by Arzoumanian et al.  (1994)
\nocite{antt94} by the statistic $\Delta_8 \equiv \log \left(
|\ddot{\nu}| t^3 /6\nu \right),$ for $t=10^8$~s.  We find
$\Delta_8 \leq 1.7$ and $ < 0.5$ for \axpa\ and \axpb, respectively.
These values place these objects clearly among the radio pulsar
population on the $\Delta_8$--$\dot{P}$ plot of Arzoumanian et
al. (1994), as was suggested for \axpb\ and 1E~1048.1$-$5937 by Heyl
\& Hernquist (1997) \nocite{hh97} on the basis of sparser, incoherent
data.  Torque noise in the accreting pulsar population would generally
predict much larger values of $\ddot{\nu}$ (\cite{bcc+97}, although
see also \cite{cbg+97}).  One caveat is that the 8~s accreting X-ray
pulsar 4U~1626$-$67 is known to be in a close binary orbit with a
low-mass companion, with orbital period 42~min (\cite{mmnw81,cha98}).
The X-ray pulsations show no Doppler shifts (\cite{lmm+88,cbg+97}),
implying that we are viewing the orbit nearly face-on.  Timing
observations of 4U~1626$-$67 over $\sim$8~yr permit phase-coherent
timing, except near one epoch where the spin-down rate abruptly
changed sign (\cite{cbg+97}).  The apparent change in spin-down rate
we have detected for \axpb\ is plausible as a change in accretion
torque.  However, if the X-rays observed for the two AXPs considered
here had an accretion origin, the orbits must both be viewed face-on
as well, which is unlikely.  Furthermore the much harder X-ray
spectrum of 4U~1626$-$67 (\cite{oop97}), which is consistent with
other accreting sources, is very different from those of any of the
known AXPs, including \axpa\ and \axpb.

An alternative model for AXPs, that they are massive white dwarfs,
also predicts regular spin down (\cite{pac90b}).  In this model, the
larger moment of inertia hence larger spin-down luminosity accounts
for the observed $L_x$.  However, if the observed X-rays have a
thermal origin, e.g. emission from hot gas heated by positrons near
the polar cap (\cite{uso93}), the implied hot spot is implausibly
small (\cite{td96a}).  Also, data obtained with the EGRET instrument
aboard the {\it Compton Gamma-Ray Observatory} show (D. Thompson,
personal communication) that the high-energy $\gamma$-ray flux from
1E~2259+586 is smaller than predicted in the white dwarf model
(\cite{uso93}).  Furthermore, an observable supernova remnant, as is
seen surrounding \axpb\ (CTB~109), is not expected for a white dwarf.
Note that the absence of a possibly associated supernova remnant is
not evidence against an AXP being a young neutron star, given the
limited observability of some remnants associated with very
young pulsars (e.g. \cite{bgl89,pkg99}).

With the stability of the rotation of \axpa\ and \axpb\ established,
the door is now open for unambiguously testing the magnetar model.  In
particular, although neither source has glitched during our
observations, which implies glitch activity lower than in some
comparably young radio pulsars (e.g. \cite{kmj+92,sl96}), future
glitches will be easily identified.  Also, for \axpa, a braking index
of 3, expected if the source is spinning down due to magnetic dipole
radiation, should be measurable in another year, although its
measurement could be complicated by timing noise, precessional effects and/or
glitches.  The periodic precession predicted by Melatos (1999) could
be clearly confirmed or ruled out in the next few years of \rxte\
monitoring, particularly if earlier observations made with other
observatories can be incorporated.

\bigskip

We thank E. Morgan, J. Lackey and G. Weinberg for help with software, A. Melatos for useful
conversations, and D. Thompson for the EGRET analysis.  VMK acknowledges
use of NASA LTSA grant number NAG5-8063.  This work was also supported
in part by NASA grant NAG5-4114 to DC.



\clearpage
\begin{figure}
\plotone{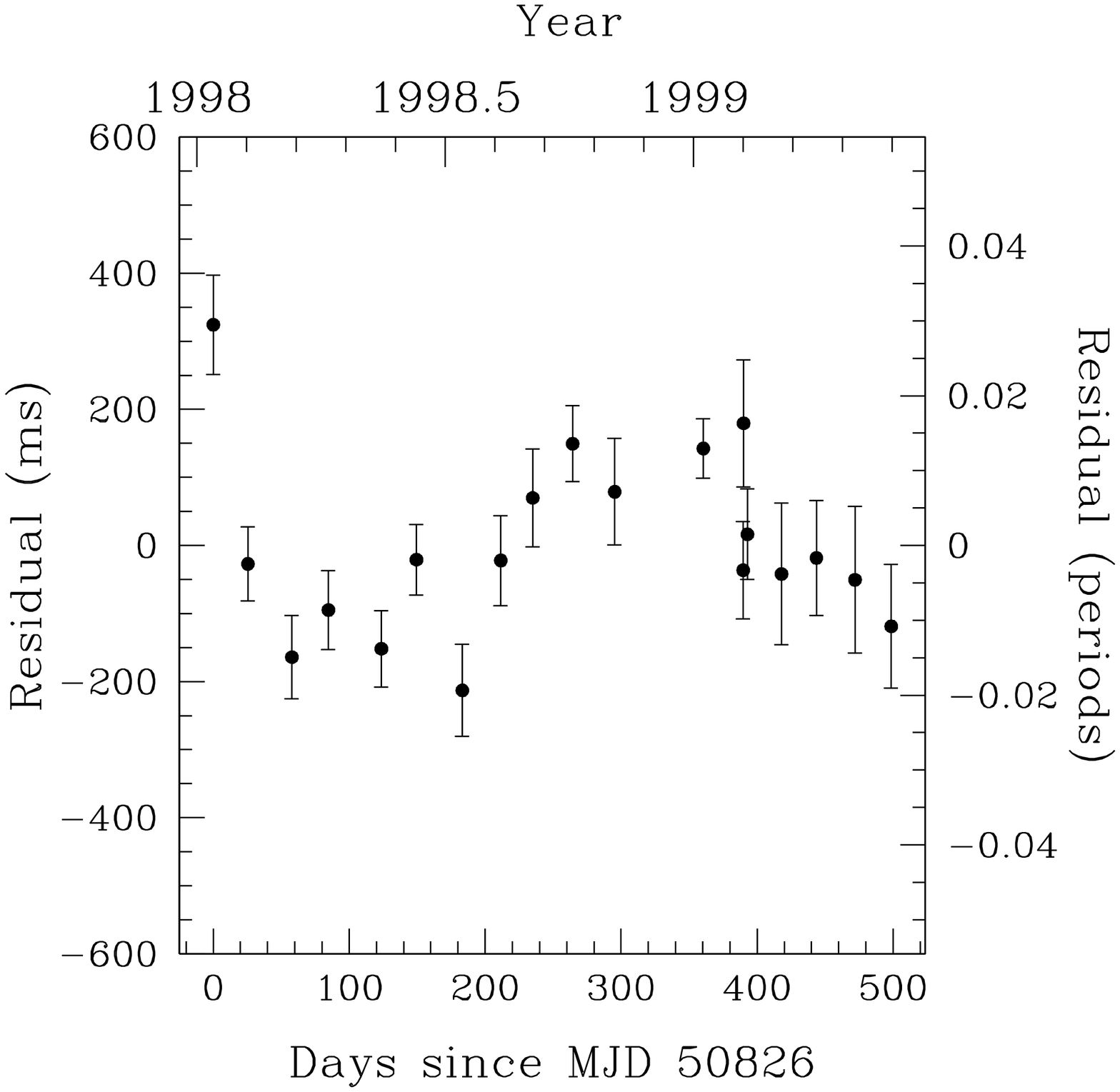}
\figcaption{Arrival time residuals for \axpa\ versus epoch.  Note
that the vertical axis represents only $\pm$5\% of the pulse period,
indicating high rotational stability.}
\label{fig:res1708}
\end{figure}

\clearpage
\begin{figure}
\plotone{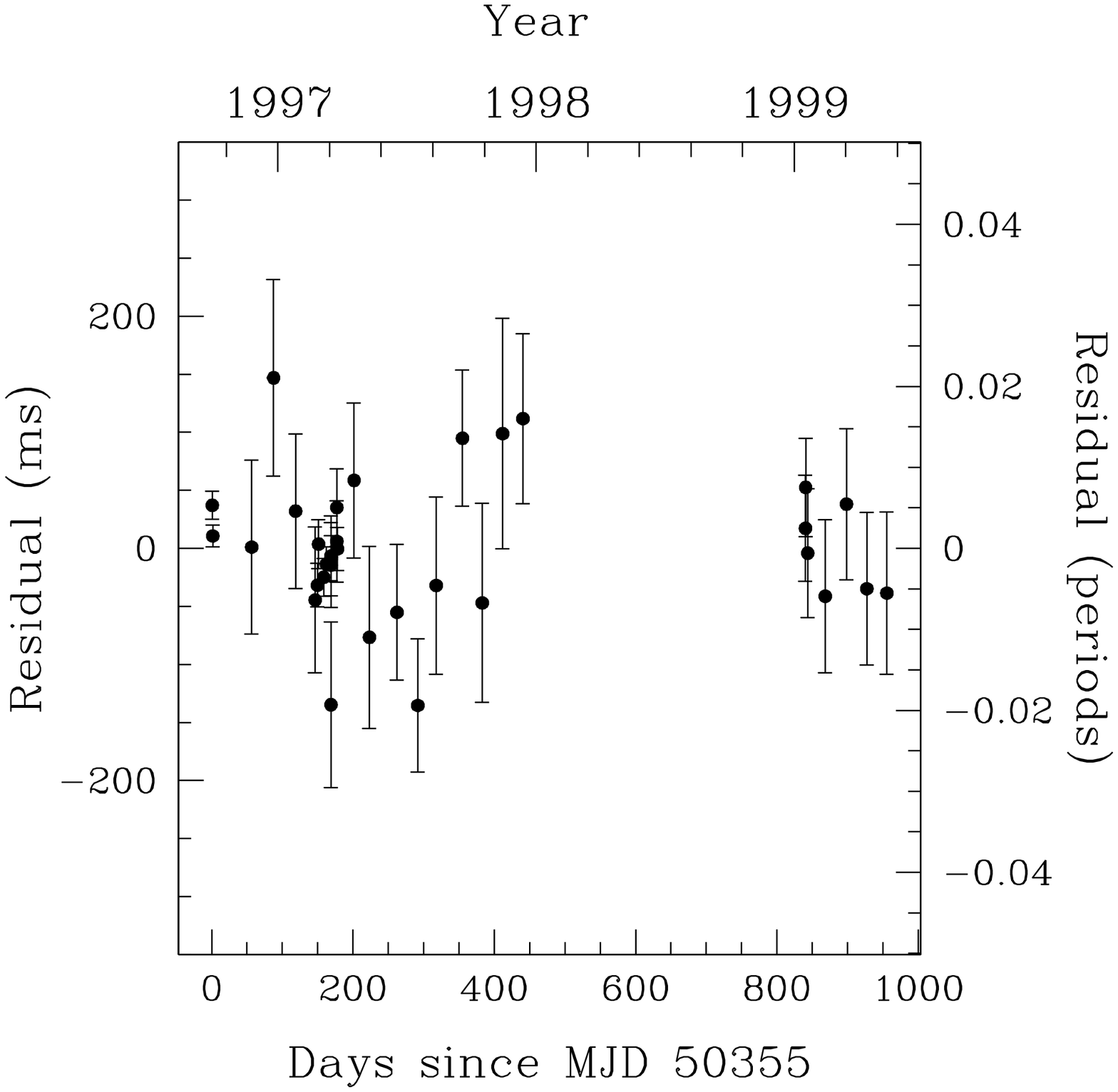}
\figcaption{Arrival time residuals for \axpb\ versus epoch.  Note
that the vertical axis represents only $\pm$5\% of the pulse period,
indicating high rotational stability.}
\label{fig:res2259}
\end{figure}

\clearpage
\begin{figure}
\plotone{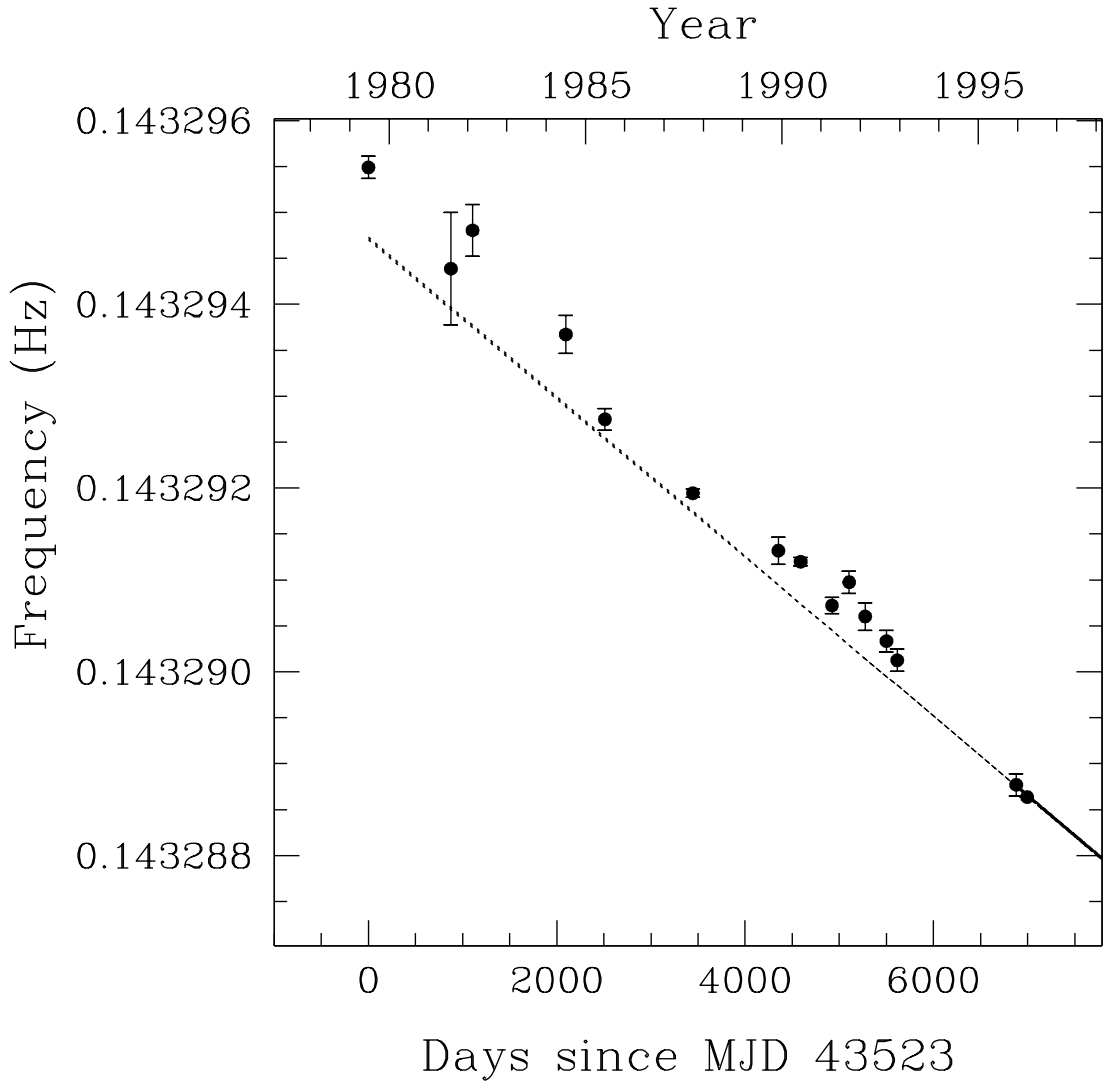}
\figcaption{Previously observed spin frequencies
for \axpb\ (see Baykal \& Swank 1996 and references therein) and the
ephemeris obtained from phase-coherent \rxte\ observations described
in this paper (see Table~\protect\ref{ta:parms}).  The epochs over
which the \rxte\ observations were made are indicated with a solid
line; previously observed frequencies that are consistent with our
spin ephemeris should fall on the dotted line.  The uncertainty in
the extrapolation is comparable the width of the dotted line.
That the previously measured frequencies fall above the line
implies that the $\dot{\nu}$ measured over 2.6~yr from the 
phase-coherent observations is not consistent with the
long-term $\dot{\nu}$.  See text for discussion.}
\label{fig:freq2259}
\end{figure}

\clearpage
\begin{deluxetable}{ccc}
\tablecaption{Parameters for Observed Anomalous X-ray Pulsars\tablenotemark{a}}
\tablehead{ \colhead{} & \colhead{\axpa} & \colhead{\axpb}}
\startdata
R.A. (J2000) & 17$^{\rm h}$ 08$^{\rm m}$ 47$^{\rm s}$ & 23$^{\rm h}$ 01$^{\rm m}$ 07$^{\rm s}$.9  \\
DEC (J2000) & $-40^{\circ}$ 08$'$ 51$''$  & +58$^{\circ}$ 52$'$ 46$''$ \\
First Observing Epoch (MJD) & 50826 & 50355 \\
Last Observing Epoch (MJD) & 51324 & 51310 \\
Total Number of Observations & 19 & 33 \\\hline
$\nu$~(Hz) &  0.0909169331(5) & 0.1432880613(2) \\
$\dot{\nu}$~($10^{-14}$ Hz s$^{-1}$) & $-$15.687(4) & $-$1.0026(7) \\
$ \ddot{\nu} $~($10^{-24}$ Hz s$^{-2}$) & [29(7)] & [2.3(1.0)] \\
$P$~(s) &  10.99905117(6)  & 6.97894850(1) \\
$\dot{P}$~($10^{-13}$)  & 189.78(5)  &  4.883(3) \\
Epoch of $\nu$ (MJD) & 51215.931 & 51195.583 \\
R.M.S. residual (ms) & 129 & 61 \\\hline
\enddata 
\tablenotetext{a}{{\it ROSAT} HRI position for \protect\axpa\ from Israel et
al. (1999).  \nocite{ics+99} Position for \protect\axpb\ from Fahlman et
al. (1982). \protect\nocite{fhr+82} Numbers in parentheses are $1\sigma$
statistical uncertainties in the last digit.  The measured $\ddot{\nu}$ for
\protect\axpa\ is marginal (see text).}
\label{ta:parms} 
\end{deluxetable}

\end{document}